\documentclass[letterpaper]{article} 
\usepackage{aaai2026}  
\usepackage{times}  
\usepackage{helvet}  
\usepackage{courier}  
\usepackage[hyphens]{url}  
\usepackage{graphicx} 
\usepackage{natbib}  
\usepackage{caption} 
\usepackage{multirow}
\usepackage{amsmath}
\usepackage{siunitx}
\usepackage{amsfonts} 

\usepackage{xcolor}
\usepackage{algorithm}
\usepackage{algorithmic}

\title{
Lina-Speech: Gated Linear Attention and Initial-State Tuning  \\ for Multi-Sample Prompting Text-To-Speech Synthesis}
\author{
    Théodor Lemerle\textsuperscript{\rm 1}, Téo Guichoux\textsuperscript{\rm 1, 2}, Axel Roebel\textsuperscript{\rm 1}, Nicolas Obin\textsuperscript{\rm 1}
}
\affiliations{
    \textsuperscript{\rm 1}Science and Technology of Music and Sound (STMS) \\ IRCAM -- CNRS -- Sorbonne Université -- Ministère de la Culture, Paris, France \\
    \textsuperscript{\rm 2} Institut des Sytèmes Intelligents et de Robotique (ISIR) \\ CNRS -- Sorbonne Université, Paris, France

}

\begin{document}

\maketitle

\begin{abstract}

Neural codec language models, built on transformer architecture, have revolutionized text-to-speech (TTS) synthesis, excelling in voice cloning by treating it as a prefix continuation task. However, their limited context length hinders their effectiveness to short speech samples. As a result, the voice cloning ability is restricted to a limited coverage and diversity of the speaker's prosody and style. Besides, adapting prosody, accent, or appropriate emotion from a short prefix remains a challenging task. Finally, the quadratic complexity of self-attention limits inference throughput. 
In this work, we introduce \textsc{Lina-Speech}, a TTS model with Gated Linear Attention (GLA) to replace standard self-attention as a principled backbone, improving inference throughput while matching state-of-the-art performance. Leveraging the stateful property of recurrent architecture, we introduce an Initial-State Tuning (IST) strategy that unlocks the possibility of multiple speech sample conditioning of arbitrary numbers and lengths and provides a comprehensive and efficient strategy for voice cloning and out-of-domain speaking style and emotion adaptation.  We demonstrate the effectiveness of this approach for controlling fine-grained characteristics such as prosody and emotion. 
Code, checkpoints, and demo are freely available: \url{https://github.com/theodorblackbird/lina-speech}
\end{abstract}

\section{Introduction}
Scaling text-to-speech \cite{tortoise} (TTS) models and data has led to drastic improvements with regard to quality, diversity, and cloning capabilities. Leveraging neural audio codecs \cite{soundstream, encodec} and next-token prediction has shown state-of-the-art results in zero-shot voice cloning, extending in-context learning abilities observed primarily on natural language to codec language. Under this setting, zero-shot voice cloning is formulated as a prompt continuation task and provides state-of-the-art performance starting with as few as 3 seconds of audio prompt. In contrast with prior works, this approach puts more pressure on the pre-training stage where large-scale speech datasets are needed in order to get sufficient in-context learning abilities and less on domain knowledge. In this direction, transformers have been the leading architecture for scalable autoregressive speech models; however, because the inherent length of speech token streams is set by the codec downsampling rate (typically ~12–75 tokens/s), the quadratic scaling of self-attention remains a key limitation. As a promising solution, several works have introduced models based on linear-attention \cite{katharopoulos2020transformers} to improve TTS models efficiency regarding long sequences.\\

In this work, we introduce \textsc{Lina-Speech}, a TTS model built on neural codec language modeling. 
\noindent The main contributions of this paper can be listed as follows:
\begin{itemize}
    \item We propose Gated Linear Attention (GLA)~\cite{gated} as a principled choice for scalable TTS, mitigating both the inference inefficiency of self-attention and the shortcomings of voice continuation by leveraging recurrent structure. In its streaming form, GLA admits a linear-RNN interpretation with a matrix-valued hidden state—the gated accumulator of key-value outer products—that serves as the model’s memory;
    \item Leveraging the persistent state in GLA, we introduce \emph{initial-state tuning} (IST) as an effective conditioning mechanism for speaker and style. IST provides multi-sample voice conditioning through optimization of the initial-state, making \textsc{Lina-Speech} a prefix-free TTS;
    \item We propose a low-rank parameterization of the initial state that stabilizes tuning across data scales and domains, while reducing embedding size and preserving output quality.
\end{itemize}
The overall architecture of \textsc{Lina-Speech} is presented in Figure \ref{fig:lina-speech}.\vspace{0.25cm}


\textsc{Lina-Speech} achieves competitive performance compared to state-of-the-art baselines in terms of naturalness and similarity. At the same time, it significantly outperforms self-attention-based codec language models in inference throughput, making it highly efficient for real-time serving.
Additionally, the experiments conducted provide empirical evidence that IST is a parameter-efficient learner for voice and speaking style cloning, especially in the challenging out-of-domain setup.

\subsection{Related Work}

\begin{figure}
    \centering
    \includegraphics[width=0.95\linewidth]{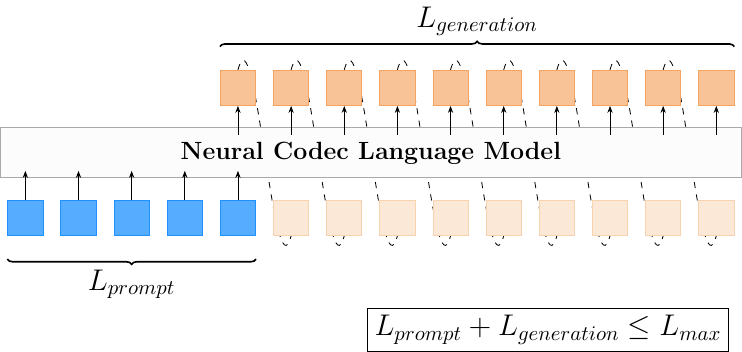}
    \caption{{\em \textbf{Voice cloning by prompt continuation imposes a trade-off between prompt length and generation length.} Neural Codec Language Models based on transformers exhibit quality degradation when generation exceeds the context length $L_{max}$, which is determined by the maximum sequence length seen during training. This creates a trade-off between prompt length and the feasible continuation length, posing a significant challenge for TTS where training samples are typically limited to under 30 seconds.}}
    \label{fig:intro}
\end{figure}

\paragraph{Principled backbones for large-scale TTS} State-of-the-art large-scale TTS models heavily relied on transformer architectures, either autoregressive (AR) \cite{valle, tortoise, lyth2024natural} or non-autoregressive (NAR) \cite{maskgit, naturalspeech, le2023voicebox}. \\

AR models have shown strong performance when trained on in-the-wild data, eliminating the need for intermediate feature representations.
Although the transformer still remains the dominant architecture for large-scale AR generative models, the attention weights learned during text-to-speech synthesis suggest that self-attention might be a suboptimal choice for this particular task \cite{lemerle24_interspeech, megatts2boostingprompting}. Indeed, as observed in previous work, transformers in the audio modality tend to focus on local information \cite{parcollet2023sumformer}, leading to weights of self-attention that are concentrated near the diagonal and a few heads of cross-attention with a strong monotonic pattern \cite{lemerle24_interspeech, tacotron2}. The use of self attention for the representation of local dependencies leads to increased computational costs and can also be seen as a lack of inductive biases towards monotonicity, which results in instabilities compared to non-autoregressive TTS models \cite{yang24l_interspeech}. Importantly, the quadratic complexity of self-attention combined with the relatively high framerate of neural audio codec prevents training with long context and is a bottleneck for inference throughput.\\

On the other hand, NAR transformers, particularly those based on diffusion or flow-matching, traditionally require either precomputed durations or an auxiliary generative model. While producing fine-grained duration annotations can be challenging for noisy, large-scale datasets, recent approaches have adopted coarser duration estimates, such as word- or sentence-level measurements \cite{yang2024simplespeech2simpleefficient}. 
Although NAR models often outperform AR models in terms of inference speed and robustness, they struggle with issues like over-smoothness \cite{yang2024simplespeech2simpleefficient, ren2022revisitingoversmoothnesstextspeech}, which leads to reduced diversity and less expressive prosody. 
Finally, recent research seeks to blend NAR and AR techniques: \cite{xin2024rallerobustcodeclanguage} introduces explicit duration modeling in an AR transformer to enhance robustness, while \cite{yang24l_interspeech} explores AR generative models for prosody and duration modeling atop a NAR flow-matching acoustic model.


\paragraph{Zero-shot TTS and Voice cloning by prompt continuation} 
Zero-shot text-to-speech (TTS) refers to the task of synthesizing speech from unseen samples during inference. 
Traditional methods include the use of speaker encoders that generate embeddings for conditioning \cite{wang2018style}. 
In contrast, large-scale TTS models leverage in-context learning capabilities with techniques like prompt continuation \cite{valle, peng2024voicecraft} and infilling strategies \cite{le2023voicebox}, and showing success using as little as 3 seconds of audio. 
These methods are robust to noisy input, such as spontaneous speech \cite{peng2024voicecraft} and in-the-wild data.\\
%

However, self-attention for TTS typically fails to extrapolate to longer transcripts than those seen during training \cite{battenberg2024robust}. As a consequence, during inference, voice cloning by continuation faces a trade-off between a long prefix, containing more information about the target speaker's voice, and a short prefix that allows the model to synthesize over a longer segment of the remaining context window (see Figure \ref{fig:intro}). 
The use of a relatively short prefix prevents the model from capturing fine details or particularities of a speaker. Typically: speech prosody, speaking style, accent, or emotions require a long observation context to fully cover the diversity and the specificity of a speaker.
Some approaches, such as Mega-TTS2 include a speaker encoder that accepts multiple samples \cite{megatts2boostingprompting}. However, they rely on speaker-labeled data, preventing training on weakly labeled data that form modern large-scale datasets.

\paragraph{Soft-prompting} 
Soft-prompting has emerged as a powerful technique for adapting pretrained language models to downstream tasks without fine-tuning their parameter set. Unlike standard forms of prompting relying on manually designed prompts, soft-prompting learns continuous vector representations that are optimized for task-specific objectives.
Prompt-Tuning \cite{liu2022ptuningv2prompttuning, xu2023parameterefficientfinetuningmethodspretrained} optimizes these embeddings directly in the input space, enabling task adaptation without modifying the model's core parameters. Prefix-Tuning \cite{li2021prefix} extends this idea by prepending learned continuous vectors, or prefixes, at every layer, effectively steering the model toward task-specific outputs. It demonstrates strong performance in NLP tasks while reducing computational overhead compared to full fine-tuning. Recent work on RWKV \cite{rwkv4, Peng2024EagleAF, istjellyfish} has demonstrated that the initial-state of its recurrent memory can be tuned for domain adaptation or instruction tuning of large language models. Since the state encodes past information without growing along the time axis, it provides a compact alternative to prompt and prefix-tuning. To the best of our knowledge, soft-prompting techniques have not yet been explored in the context of speech synthesis.


\section{Preliminaries}

Given an input $\mathbf{X} \in \mathbb{R}^{N \times d}$ self-attention for autoregressive modeling uses the following three linear projections: the query matrix \(\mathbf{Q} \in \mathbb{R}^{N \times d_k}\), the key matrix \(\mathbf{K} \in \mathbb{R}^{N \times d_k}\), the value matrix \(\mathbf{V} \in \mathbb{R}^{N \times d_v}\),  and a causal mask \(\mathbf{M}_{i,j} = \mathbf{1}_{i<j}\) \( \mathbf{M} \in \{0,1\}^{N \times N}\). The parallel form of attention is defined as:

\begin{equation*}\label{eq:att}
\text{Attention}(\mathbf{Q}, \mathbf{K}, \mathbf{V}) = \text{Softmax}\left(\frac{\mathbf{Q}\mathbf{K}^\top}{\sqrt{d}} \odot \mathbf{M}\right)\mathbf{V},
\end{equation*}

where \( \odot \) denotes element-wise multiplication, and admits the sequential form,
\begin{align*}
     \text{Attention}(\mathbf{Q}, \mathbf{K}, \mathbf{V})_t &= \frac{\sum_{i=1}^t exp(\mathbf{q_t k_i^\top}) \mathbf{v_i}}{\sum_{i=1}^t exp(\mathbf{q_t k_i^\top})}.
\end{align*}
during inference.

\subsection{Linear Attention}

\cite{katharopoulos2020transformers} proposed to replace the softmax in self-attention with a general kernel function $k$ and its associated feature map \(\phi\). This approach, known as linear attention, can be expressed as:

\begin{align}
     \text{LinearAttention}(\mathbf{Q}, \mathbf{K}, \mathbf{V})_t &= \frac{\sum_{i=1}^t \phi(\mathbf{q_t}) \phi(\mathbf{k_i})^\top \mathbf{v_i}}{\sum_{i=1}^t \phi(\mathbf{q_t}) \phi(\mathbf{k_i})^\top}.
     \label{eq:linattlong}
\end{align}

\noindent Denote \begin{align*} \mathbf{S_t} = \sum_{i=1}^t \phi(\mathbf{k_i})^\top \mathbf{v_i},\quad \mathbf{z_t}=\sum_{i=1}^t \phi(\mathbf{k_i})^\top, \mathbf{o_t} = \frac{\mathbf{\phi({q_t}}) \mathbf{S_t}}{\mathbf{\phi(q_t)z_t}} \end{align*} 
Eq. \ref{eq:linattlong} can be reformulated into a recursive form through the update rule:
\begin{equation}\label{eq:linattsimp}
\begin{aligned}
\mathbf{S_t} &= \mathbf{S_{t-1}} + \phi(\mathbf{k_t})^\top \mathbf{v_t},\\
\mathbf{z_t} &= \mathbf{z_{t-1}} + \phi(\mathbf{k_t})^\top,
\mathbf{o_t} &= \frac{\phi(\mathbf{q_t})\,\mathbf{S_t}}{\phi(\mathbf{q_t})\,\mathbf{z_t}}.
\end{aligned}
\end{equation}
revealing that it essentially functions as a recurrent neural network with a matrix-valued state. \\

The choice of $\phi$ being the linear kernel ($\phi = Id$) has been a popular line of research \cite{Peng2024EagleAF, gated, retnet}. Furthermore, it has been observed that in practice the normalization term can be omitted, thus simplifying Eq. \ref{eq:linattsimp} into:

\begin{equation}\label{eq:linatt}
    \mathbf{S_t} = \mathbf{S_{t-1}} + \mathbf{k_t}^\top \mathbf{v_t}, \quad \mathbf{o_t} = \mathbf{q_t} \mathbf{S_t},
\end{equation}
\noindent where: \( \mathbf{S_t} \) acts as a constant-size kv-cache in traditional self-attention transformer.

\subsection{Gated Linear Attention (GLA)}

While linear attention provides a constant memory footprint and achieves linear time complexity during inference, its parallel form remains constrained by quadratic time complexity, and the recurrent form poses challenges for efficient training on modern hardware. Recent advances in linear-complexity language models—such as RWKV-6 \cite{rwkv4, Peng2024EagleAF}, GLA \cite{gated}, and Mamba \cite{mambassm, dao2024transformersssmsgeneralizedmodels} demonstrate that introducing data-dependent gating mechanism \cite{retnet, rwkv4} substantially closes the performance gap with self-attention transformers. Additionally, various techniques have been proposed to enhance hardware-efficiency for linear-scaling language models, including the prefix-sum algorithm \cite{mambassm, katsch2023gateloop} and chunk-wise computation \cite{gated, dao2024transformersssmsgeneralizedmodels, retnet}.\\

\begin{figure*}[h!]
    \centering
    \includegraphics[width=0.85\linewidth, trim={0 0.0cm 0 -0.25cm}]{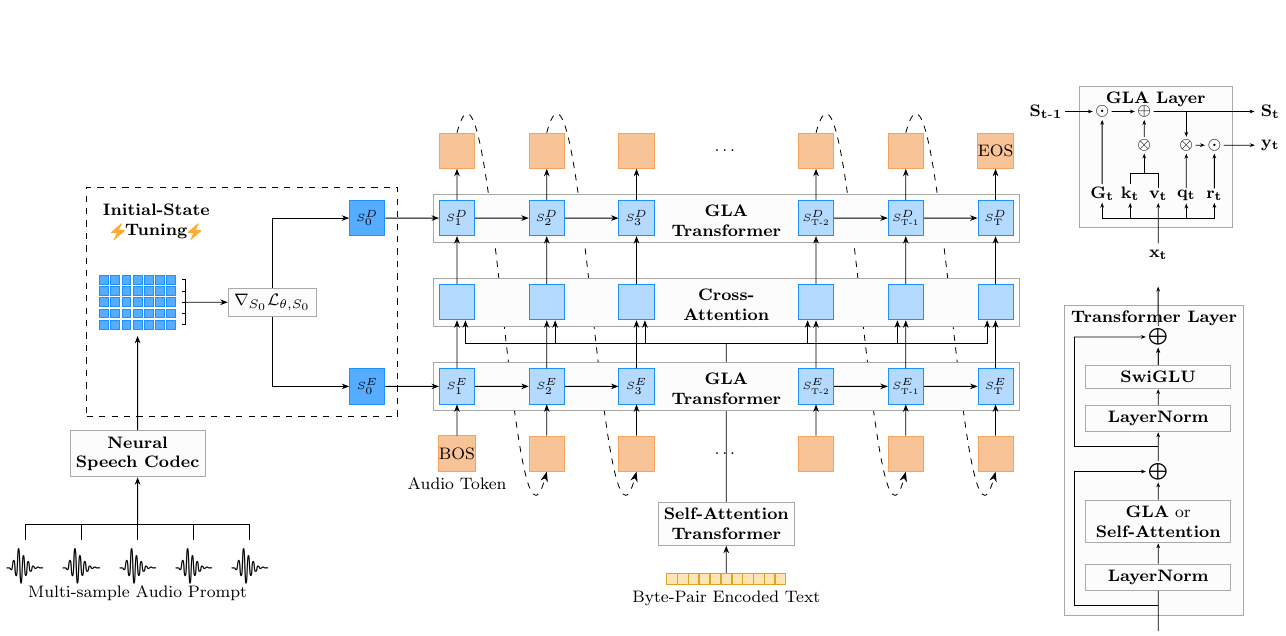}
    \caption{{\em \textsc{Lina-Speech} model. $S^E_t$ and $S^D_t$ are encoder and decoder states at time-step $t$ respectively. These states consist of one matrix per GLA layer and per head. For $t=0$, they default to $\mathbf{0}$ but can be tuned efficiently on a specific speaker or style. Initial-state tuning consists of replacing the initial 0 by means of an initial state that is learned using a soft prompt while freezing the models parameters $\theta$.}}
    \label{fig:lina-speech}
\end{figure*}

For these reasons, Gated Linear Attention \cite{gated} (GLA) comes with a data-dependent structured gating mechanism, resulting in the following update rule:
\begin{equation}
    \mathbf{S_t} = \mathbf{G_t} \odot \mathbf{S_{t-1}} + \mathbf{k_t^\top} \mathbf{v_t},  \quad \mathbf{o_t} = \mathbf{q_t} \mathbf{S_t},
    \label{eq:glaeq}
\end{equation}
where: \(\mathbf{G_t=\alpha_t^\top1}\) is a decay matrix that modulates the contribution of past states.
\paragraph{Performance} GLA achieved state-of-the-art results at linear-complexity language modeling, even matching or surpassing transformer models for some tasks at large scale.
\paragraph{Efficiency} GLA admits hardware-efficient implementation \cite{gated} by imposing some structure to the gating term $\mathbf{G_t}$ and leveraging chunk-wise calculation of Eq. (\ref{eq:glaeq}). It enables higher inference throughput compared to a similar size transformer on long sequences when doing batch inference. Its recurrent nature and constant memory footprint make it an attractive option for tasks like audio modeling, streaming, or on-device applications.
\paragraph{Principled choice for scalable TTS} While these linear complexity language models are known to underperform on recall-intensive tasks \cite{arora2024simplelinearattentionlanguage}, we hypothesize that they could mitigate the inefficiency of self-attention in domains like speech modeling \cite{ parcollet2023sumformer, lemerle24_interspeech, megatts2boostingprompting}, where it appears to be less critical or even unnecessary.


\section{Method}

\textsc{Lina-Speech} is an autoregressive generative model \(p_\theta\) designed to approximate the distribution of neural audio codec token sequence, denoted as \( c \), conditioned on text input \( x \), with \(\theta\) representing the model parameters, that is:

\begin{equation}
    p_{\theta}(c|x) = \prod_{t=1}^T p_{\theta}(c_t | c_{<t}, x).
\end{equation}

\subsection{Model architecture and inference}
\noindent The general architecture of \textsc{Lina-Speech} is presented in Figure \ref{fig:lina-speech}.

\paragraph{Model architecture} 
 The text encoder is a non-causal transformer encoder with self-attention as time-mixing operator and SwiGLU \cite{Shazeer2020GLUVI} as a feed-forward network. It employs RoPE positional encoding \cite{su2024roformer}. The acoustic model includes both an audio encoder and a decoder, featuring a causal transformer architecture with GLA as a time-mixing operator, SwiGLU \cite{Shazeer2020GLUVI} as a feed-forward network, and no positional encoding.

The decoder takes input from the audio encoder and a cross-attention layer between the text and audio encoder outputs. To improve robustness, we used the position-aware cross-attention from \cite{lemerle24_interspeech}, and replaced sinusoidal positional encoding with convolutional positional encoding for enhanced training stability.

\paragraph{Inference} We use top-$k$ sampling with $k=100$ and treat the \textbf{EOS} token as an additional token in the audio codebook.

\subsection{Initial-state tuning}

We have seen that Gated Linear Attention achieves linear complexity by replacing the expanding key-value cache of traditional transformers with a constant-sized memory, represented by the matrix-valued state $\mathbf{S_t}$ in Eq. (\ref{eq:glaeq}). During inference, the initial states are initialized by default to zeros, i.e., $\mathbf{S_0} = \mathbf{0}$. This memory can be subject to soft-prompting by treating them as learnable parameters while freezing the model parameters. We refer to this strategy as initial-state tuning (IST) and it can be formalized as follows:

\noindent Let \(S^{(i)}_{0}(\phi)\) denote the learnable initial state of layer \(i\), and let
\(S_{0}(\phi)=\{S^{(i)}_{0}(\phi)\}_{i=1}^{L}\).
We write the TTS backbone as \(f(x;\,\theta, S_{0}(\phi))\), where
\(\theta\) are pretrained weights kept frozen.
Given paired text-speech samples \((x,y)\sim\mathcal{D}\) of the target speaker's voice, we optimize only \(\phi\):
\[
\min_{\phi}\;
\mathbb{E}_{(x,y)\sim\mathcal{D}}
\Big[\mathrm{CE}\big(f(x;\,\theta, S_{0}(\phi)),\, y\big)\Big],
\qquad \text{with } \frac{\partial \mathcal{L}}{\partial \theta}=0.
\]
In practice, we found that a low-rank matrix representation of the Initial States, $S_{0}(\phi)$, improved performance. This aspect is further discussed in Section \ref{sec:exp3}.

\noindent 

\paragraph{IST for Prefix-Free and Multi-Sample Prompting TTS} Voice continuation relies on using audio and text references as a prefix, which reduces the remaining available context length. In contrast, IST relies solely on the initial-state, enabling generation up to the maximum length observed during training. Because the resulting state is text-agnostic, it prevents semantic leaks from a particular text prompt.
We show that this approach is particularly well-suited for voice cloning and adaptation using multiple samples without requiring any architectural or training modifications. 
In practice, initializing the model's recurrent state with an observation of arbitrary length and number of segments provides a richer context window. This method enables the model to more accurately capture the distribution and diversity of speech prosody, addressing a key challenge in voice cloning either with speaker embedding or voice continuation. \\
%

Experimental evaluation reported in Section \ref{sec:exp3} provide empirical evidence that IST is an efficient strategy for zero-shot voice-cloning and speaking style adaptation, among some other empirical properties such as: \textbf{fast tuning}, \textbf{efficient low-rank approximation} of the initial-state, and \textbf{low-sensitivity to tuning parameters}.

\section{Experiments}

Three experiments were conducted to assess the performance of the proposed \textsc{Lina-Speech} architecture for TTS, by comparison with existing baselines, and by providing specific further experiments to address the efficiency of Gated-Linear Attention (GLA) and Initial-State Tuning (IST). The remainder of this section describes the three experiments, the general experimental setup, the metrics used, and the baselines selected for the comparison.

\subsection{Experiment \#1: Zero-Shot Voice Cloning} As a first experiment, we conducted an evaluation of \textsc{Lina-Speech} on a zero-shot voice cloning task, using voice continuation and initial-state tuning, with comparison to the baselines. 
In order to assess the zero-shot voice-cloning ability both on in- and out-of-domain datasets, we conducted two series of evaluations.
%
For the in-domain setup, we evaluated on the two test splits of LibriTTS. For the out-of-domain setup, we evaluated on the  Expresso dataset \cite{nguyen2023expresso}, which consists of studio-quality recordings of 4 speakers labeled with different emotions or styles (happy, sad, whisper ...). 
As a baseline for this experiment, we fine-tuned Parler-TTS from the official repository on the Expresso dataset. 
%

\subsection{Experiment \#2: Focus Study on GLA vs. Self-Attention}
A second experiment was conducted with a particular focus on the properties of the attention mechanisms used in \textsc{Lina-Speech}. In particular, a comparison of Self-Attention (SA) and Gated Linear Attention (GLA) within the same \textsc{Lina-Speech} architecture is provided.
Firstly, SA and GLA were compared in terms of inference speed.
Then, they were additionally compared on the Zero-Shot Voice-Cloning task described in the previous experiment.

\subsection{Experiment \#3: Focus Study on Initial-State Tuning}

A third experiment was conducted with a particular focus on the properties of the IST in \textsc{Lina-Speech}.

\subsection{Experimental setup}

\paragraph{Datasets} We trained \textsc{Lina-Speech} on a publicly available English subset of MLS\footnote{\url{parler-tts/mls_eng_10k}} \cite{lacombe-etal-2024-parler-tts} which consists of 10k hours of librivox recordings. We do not use the provided transcription and rather use the Automatic Speech Recognition (ASR) model NeMo\footnote{\url{https://catalog.ngc.nvidia.com/orgs/nvidia/teams/nemo/models/stt_en_fastconformer_hybrid_large_pc}{stt\_en\_fastconformer\_hybrid\_large\_pc}}. We also added both LibriTTS \cite{zen2019libritts} and its restored version LibriTTS-R \cite{koizumi2023libritts} with their normalized transcripts. We used WavTokenizer \cite{ji2024wavtokenizer}\footnote{\url{https://huggingface.co/novateur/WavTokenizer-medium-speech-75token/tree/main}{WavTokenizer-medium-speech-75token}}  
as a neural audio codec that encodes speech at a rate of 75 token/s, with a codebook size of 4096 \cite{koizumi2023libritts}. For text representation, we trained a byte-pair encoding tokenizer with a vocabulary size of 256 on the lower-cased transcripts from LibriTTS.

\paragraph{Training and Inference} The main model is trained for next-token prediction with cross-entropy loss for 500k steps with a batch size of approximately 100k tokens ($\approx22$~min of speech). We use AdamW optimizer with a learning rate of \num{2e-4}, a cosine learning rate schedule with linear warmup for the first 1k steps, a weight decay of $0.1$ and gradient clipping of $1.$  We group samples of similar lengths within 10 buckets in order to avoid padding. We rely on the official hardware-efficient implementation of GLA provided in the flash-linear-attention repository\cite{yang2024fla}.

\paragraph{Objective metrics} We measure word error rate (WER) and character error rate (CER) using the same ASR model from NeMo as for speech transcription. We also measured speaker similarity (Sim-O) as the cosine similarity of WavLM \cite{chen2022wavlm} embedding of target and synthesized speech using a pretrained checkpoint\footnote{\texttt{\url{wavlm-base-plus-sv}}}.


\paragraph{Subjective metrics} We conducted a subjective experiment using Mean Opinion Score (MOS) to rate the perceived naturalness (N-MOS) and similarity to the target speaker (S-MOS) via the platform Prolific.
165 subjects participated in the experiment, each subject rated 20 speech stimuli randomly drawn from real-speech,  \textsc{Lina-Speech} and the baselines generated speech.
We applied several filters to assess the qualification of the subjects, to reject those who do not fulfill the necessary conditions to be considered qualified for the evaluation.
The list of exclusion conditions comprises: non-native English speaker, rate below 3 any real speech sample on the N-MOS, time spent to complete the experiment is below 3m 30s, the mean MOS of the subject deviates from the overall mean of all subjects by more than two standard deviations, as proposed by \cite{CLAM-TTS}.
A subject was considered not qualified if at least one of the conditions was not fulfilled.
Applying these filters, the qualification rate of the subjects was about 76\%. We rejected 39 subjects from a total of 165, so the total of qualified subjects was 126 whose ratings were further used for analysis.

\begin{table*}[ht!]
\small
    \centering
     \caption{{\em \textbf{Zero-shot voice-cloning experiment} conducted on in-domain (LibriTTS test clean split) and out-of-domain (Expresso) datasets. The objective evaluation includes: Word Error Rate (WER), Character Error Rate (CER), and cosine similarity to the reference speaker (Sim-O). The subjective evaluation includes: MOS for naturalness (N-MOS) and MOS for similarity to the reference speaker (S-MOS). The results obtained for the subjective evaluation are reported along with their 95\% confidence interval. \textsc{Lina-Speech} presents the highest scores both in terms of naturalness and similarity to the reference speaker. The number of parameters for each model is reported in \#Params.  \vspace{0.15cm}}}
    \begin{tabular}{c | l | l | c c c | c c | c}
         \multicolumn{1}{c}{Dataset}& \multicolumn{1}{c|}{Model} & \multicolumn{1}{c|}{Method} & \multicolumn{3}{c|}{Obj. Eval.} & \multicolumn{2}{c|}{Subj. Eval.} & \#Params. \\
         \cline{4-8}
        & &  &{\scriptsize CER$\downarrow$ }&{\scriptsize WER$\downarrow$}&{\scriptsize Sim-O$\uparrow$}&{\scriptsize N-MOS$\uparrow$}& {\scriptsize S-MOS$\uparrow$}&   \\
        \hline
         \multirow{7}{*}{\shortstack{\textsc{LibriTTS}\\(in-domain)}}
         &Ground Truth & - &1.5\% & 4.5\% & - & $4.41 \pm 0.14$  & $4.31 \pm 0.17$    &  -\\
         \cline{2-9}
         &\textsc{StyleTTS2}  & Sp. Enc. &\pmb{0.8}\% & \pmb{3.2}\% & 0.89 & $4.02 \pm 0.16$ & $3.95 \pm 0.22$  & 148M  \\
         &\textsc{XTTS} v2 & Sp. Enc. &2.5\% & {5.5\%} & {0.93} & $3.62 \pm 0.20$ & $3.23 \pm 0.24$  & 443M \\
         &\textsc{VoiceCraft} & Pr. Cont. &2.8\% & {7.5\%} & {0.94} & $3.73 \pm 0.24$ & $3.57 \pm 0.23$  & 830M \\
         &\textsc{CosyVoice2} & Pr. Cont. &1.1\% & {3.8\%} & \pmb{0.95} & $3.89 \pm 0.19$ & $3.98 \pm 0.18$  & 500M \\
         &\textsc{Lina-Speech} (ours) & Pr. Cont. &{2.8\%} & 6.9\% & {0.93} & ${4.14\pm0.20}$ & ${4.07\pm 0.18}$  & 311M \\
         &\textsc{Lina-Speech} (ours) & IST &{2.8\%} & 6.5\% & {0.93} & $\pmb{4.16\pm0.19}$ & $\pmb{4.14\pm 0.17}$  & 311M \\
    
         \hline
         \hline

%
%



        \multirow{7}{*}{\shortstack{\textsc{Expresso}\\(out-of-domain)}}&Ground Truth & - & 1.6\% & 5.1\% & - & $4.59 \pm 0.24$  & $4.34 \pm 0.22$  & - \\
        \cline{2-9}
         &\textsc{Parler-TTS} (FT)& Nat. Lang. Pr. & 2.6\% & 4.4\% & {0.88} & ${3.59 \pm 0.29}$ & $3.41 \pm 0.28$  & 674M \\
        &\textsc{XTTS} v2 & Sp. Enc. & 1.0\% & 2.7\% &0.85 &  $3.64\pm0.26$ & $3.09\pm0.27$  & 443M \\

         &\textsc{VoiceCraft} & Pr. Cont. & 3.5\% & 5.4\% &0.85 &  $3.54\pm0.20$ & $3.14\pm0.24$  & 830M \\
         &\textsc{CosyVoice2} & Pr. Cont. & \pmb{0.4}\% & \pmb{1.7}\% & \pmb{0.89} &  $
         {3.85\pm0.26}$ & $3.25\pm0.27$  & 500M \\
         &\textsc{Lina-Speech} (ours)& Pr. Cont. & ${1.1\%}$ & ${3.1\%}$ & 0.82 & $3.79 \pm 0.24$ & $3.11 \pm 0.27$ & 311M \\
         &\textsc{Lina-Speech} (ours) & IST  & ${1.3\%}$ & ${3.2\%}$ & ${0.86}$ & \pmb{${3.94 \pm 0.20}$} & $\pmb{3.63\pm0.28}$ & 311M\\
        \hline

    \end{tabular}
    \label{tab:IST_eval}
\end{table*}

\subsection{Baselines}
\noindent The baselines used for comparison include: 
\begin{itemize}
    \item The TTS enhanced version of VoiceCraft \cite{peng2024voicecraft}, a decoder-only transformer trained on GigaSpeech and Libri-light, which includes an EnCodec model specifically trained for speech.
    \item StyleTTS2 \cite{li2024styletts} an end-to-end TTS model that leverages latent diffusion for style modeling. We evaluate it only on LibriTTS as we have found it unable to adapt to highly expressive data.
    \item Parler-TTS \cite{lacombe-etal-2024-parler-tts}, is a series of reproduction of \cite{lyth2024natural} that allows synthesis controlled by textual description of the voice. Interestingly, this reproduction differs from the original paper by separating text and audio sequence and employing cross-attention between the two modalities instead of self-attention on the concatenation of both, making the architecture closer to \textsc{Lina-Speech}. They leverage DAC \cite{kumar2024high} as audio codec. We used a fine-tuned checkpoint on \textsc{Expresso} and evaluated it on this dataset only.
    \item XTTS v2 \cite{casanova24_interspeech}, a large-scale multilingual TTS model that extends on the architecture of Tortoise \cite{tortoise}.
    \item CosyVoice2 \cite{du2024cosyvoice}, a recently introduced large-scale Neural Codec LM combined with a flow-matching decoder, which includes many improvements related to semantic modeling and is built on top of a text language model. 
\end{itemize}

The choices of the baselines provides a strong benchmark of existing TTS models with a variety of strategies for zero-shot voice-cloning (see Fig. \ref{tab:IST_eval}): speaker embedding (StyleTTS2 and XTTS v2), prompt continuation (VoiceCraft, CosyVoice2, and \textsc{Lina-Speech}), natural language prompting (Parler-TTS), and initial-state tuning (\textsc{Lina-Speech}).

\section{Results}
\label{sec:results}

This section presents the results obtained for the three experiments.

\subsection{Experiment \#1: Zero-Shot Voice-Cloning}

Table \ref{tab:IST_eval} presents the results obtained for in- and out-of-domain datasets with \textsc{Lina-Speech} and the other baselines.
%
On the \textsc{LibriTTS} dataset, \textsc{Lina-Speech} performs competitively with existing models. When using Initial-State Tuning, \textsc{Lina-Speech} shows comparable performance to other leading models in terms of both objective and subjective metrics. Notably, \textsc{Lina-Speech} achieves the highest MOS scores for naturalness and speaker similarity, outperforming several models with significantly larger parameter sizes.
%
%
On the \textsc{Expresso} dataset, \textsc{Lina-Speech} continues to demonstrate strong performance, particularly in speaker similarity. While other models, such as \textsc{CosyVoice2}, show better results in objective evaluations, \textsc{Lina-Speech} (Initial-State Tuning) achieves notable improvements in subjective evaluations, particularly in speaker similarity.


\paragraph{Prompt Continuation vs. Initial-State Tuning}
\textsc{Lina-Speech} with IST consistently enhances speaker similarity (S-MOS) compared to \textsc{Lina-Speech} with the standard voice continuation method, both on in- and out-of-domain datasets. The improvement on \textsc{Expresso} is particularly noteworthy, as \textsc{Lina-Speech} was trained exclusively on LibriVox recordings, while other baselines include models that have undergone fine-tuning or large-scale pretraining. 

\subsection{Experiment \#2: GLA vs. Self-Attention}


\begin{figure}[h!]
    \centering
    \includegraphics[width=1.0\linewidth]{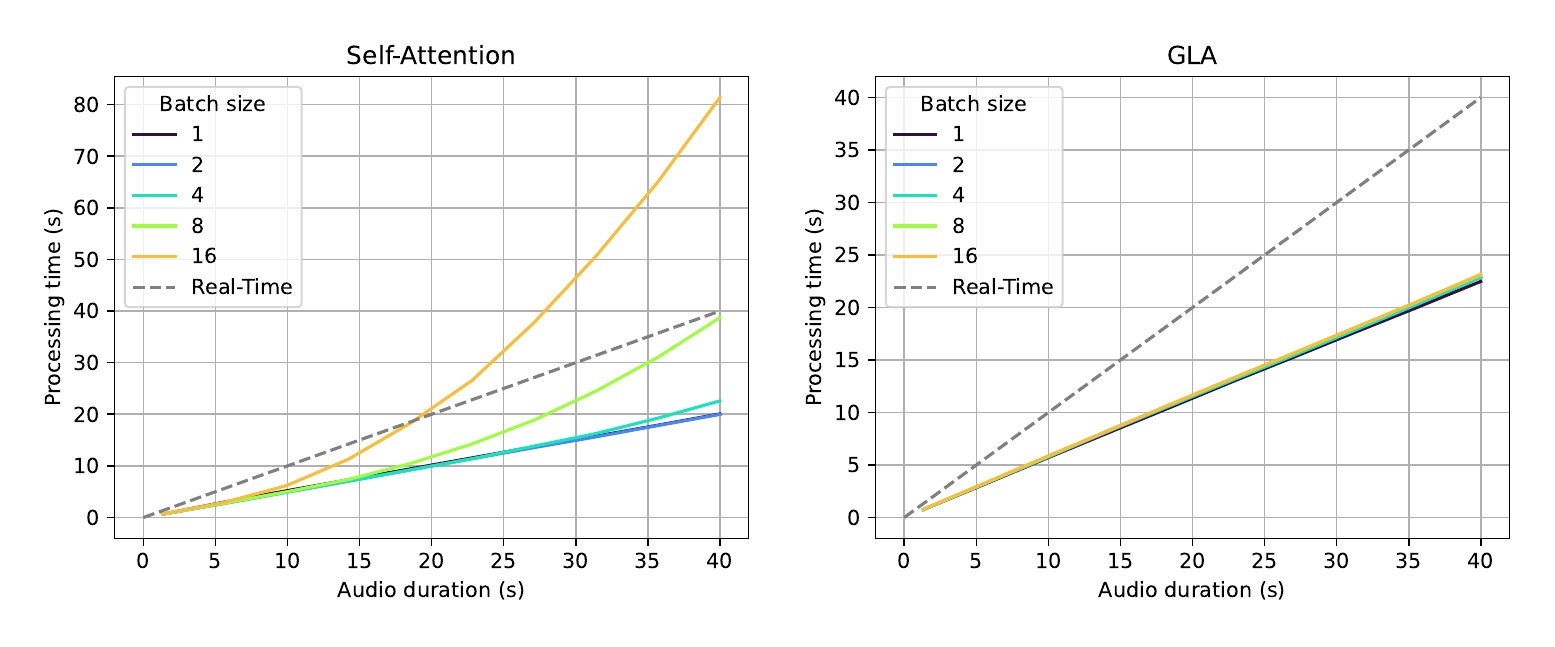}
    \caption{{\em \textbf{Inference speed comparison between self-attention and gated linear attention.} The inference speed was measured on a RTX4090 for varying batch sizes. We compared Lina-Speech against a Self-Attention equivalent model. While self-attention is slightly faster for small batch sizes, Lina-Speech benefits from a much higher inference throughput.}}
    \label{fig:lina-abblation}
\end{figure}


\paragraph{Efficient inference}
The linear nature of GLA greatly improves inference throughput over self-attention for batched generation and long sequences (see Fig. (\ref{fig:lina-abblation})) and presents slightly better performance (see Tab. \ref{tab:lina-abblation}).
In particular, the inference throughput of GLA is below real-time inference regardless of the batch size, making it an attractive option for efficient low-latency serving.
Moreover, once tuned with IST the state can be reused across different generations, which makes the tuning process amortizable, unlike voice continuation. Indeed, prefix continuation incurs a constant prefill cost and adequate padding for batched generation. In contrast, since IST leaves the model weights unchanged so that a single model can generate in parallel from different initial states for the same computational cost as unconditional generation.

\begin{table}[h!]
\centering
\caption{{\em GLA Ablation study. We report test perplexity as ppl and evaluate on LibriTTS test clean split.}}
\begin{tabular}{l | c c | c | c}
\hline
& \scriptsize CER $\downarrow$ & \scriptsize WER $\downarrow$ & \scriptsize Sim-O $\uparrow$ & \scriptsize ppl $\downarrow$ \\
\hline
Lina-Speech w. GLA & \textbf{2.8} & \textbf{6.9} & 0.93 & \textbf{49.6} \\
Lina-Speech w. SA & 3.2 & 7.8 & 0.93 & 50.4 \\
\hline
\end{tabular}
\label{tab:lina-abblation}
\end{table}


\subsection{Experiment \#3: Initial-State Tuning}
\label{sec:exp3}

\begin{figure}[h!]
    \centering
    \includegraphics[width=0.75\linewidth]{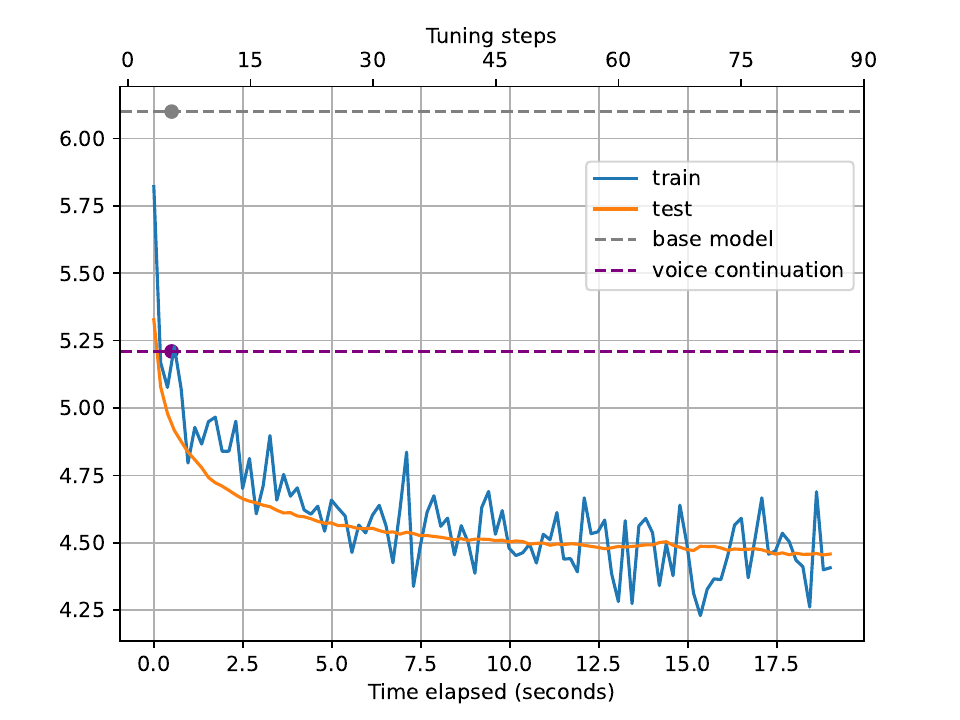}
    \caption{{\em \textbf{Initial-State Tuning convergence speed}. IST converges rapidly, typically within 100 steps, with an average runtime of under 20s on an RTX 4090. Example shown for speaker \texttt{ex01} with the emotion "sad" from the \textsc{Expresso} dataset. We report training and test losses. We also reported the loss averaged over 16 different prompts (voice continuation) and unconditioned (base model) for comparison.}}
    \label{fig:ist_tuning}
\end{figure}

\paragraph{Efficient tuning} The method typically achieves convergence within 100 steps (see Fig. (\ref{fig:ist_tuning})), with a runtime ranging from 5s to 20s on an RTX 4090 for a target speaker or style, providing from 45s to 20min of audio.

\begin{figure}[h!]
\hspace{-1cm}
    \includegraphics[width=1.2\linewidth]{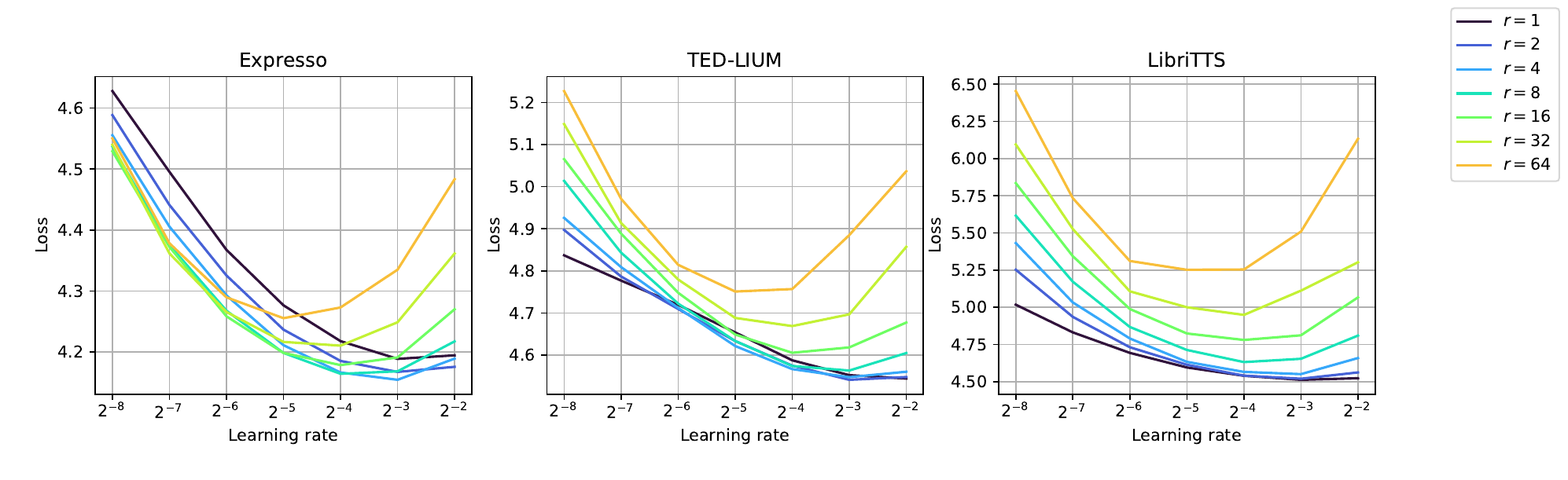}
    \vspace{-0.5cm}
    \caption{{\em \textbf{Impact of the rank and learning rate for initial-state tuning} on the test loss, from left to right on Expresso, TED-LIUM and LibriTTS datasets. For each dataset we report the best test loss averaged over 20 random speakers/styles. In particular, initial-state parameterized as a rank-one matrix performs best on TED-LIUM \cite{hernandez2018ted} and LibriTTS and is close to the best rank on Expresso. Notably, the optimal learning rate does not vary across datasets.}}
    \label{fig:rank}
\end{figure}

\paragraph{Low-rank initial-state} 
We successfully experimented with a special variant of IST, where $\mathbf{S_0}$ is represented as a low-rank matrix. 
In practice we have found that the state matrix can be parameterized as a rank-1 matrix (that is $\mathbf{S_{0}} = \mathbf{k_{0}^Tv_{0}}$), reducing the parameters set to a pair of vectors per head and per layer, while maintaining across all datasets near optimal performance (see Fig. (\ref{fig:rank})), and demonstrated in all cases, better performance than a full-rank initial-state matrix.

 \paragraph{Low sensitivity to tuning parameters} We have found initial-state tuning to be remarkably stable across different datasets with various backgrounds, recording conditions and sizes (see Fig. (\ref{fig:rank})), such as audiobooks \cite{zen2019libritts}, high-quality expressive speech \cite{nguyen2023expresso} or recorded conferences \cite{hernandez2018ted}. In practice, we do not need to tune the learning parameters for each speaker nor do we use early stopping as we have found the low-rank parametrization to play a crucial role in regularization. This contrasts with fine-tuning which typically needs supervision in order to avoid catastrophic forgetting, especially when the amount of data is low (\textit{e.g.,} few minutes). In practice, and in all the experiments below, we tune $\mathbf{S_0}$ using AdamW, a learning rate of $2^{-3}$, a batch size of 8 for a maximum of 100 steps over the samples per speaker or style. 

\subsection{Summary}

The main empirical findings of the three experiments presented in Section \ref{sec:results} can be summarized as follows.\\
 
First, \textsc{Lina-Speech} achieves strong performance, particularly in naturalness and speaker similarity, while maintaining a moderate model size. In particular, it remains competitive in both objective and subjective evaluations with comparable or higher performance when compared to larger models such as \textsc{VoiceCraft} and \textsc{CosyVoice2}.
It also presents a constant memory footprint and a much higher inference throughput compared to the standard self-attention.

Second, IST is proven to be efficient for zero-shot voice and speaking style cloning, in particular to out-of-domain speaking styles or emotions. It outperforms all the existing strategies used for voice-cloning, either with speaker embeddings or prompt continuation.
We observed that the multi-sample prompting allows to reproduce more finely the diversity and the specificity of the speaking style of a speaker. 
Furthermore, IST is simple and fast to tune, has an efficient low-rank approximation, and has low-sensitivity to tuning parameters.\\

%
%

%
In conclusion, \textsc{Lina-Speech} presents comparable performance to the other LM TTS models in terms of objective metrics and demonstrates significant improvements in terms of subjective metrics, while having much less parameters than the other models. We note a slight degradation in naturalness that co-occurs with the increase in similarity with initial-state tuning. We associate this with the increasing difficulty related to the highly expressive data that composes \textsc{Expresso} and the fact that our training data is less diverse than other baselines, as corroborated by the lowest similarity in prompt continuation among all.



\section{Limitations and future work}
\paragraph{Audio Codec} We found that WavTokenizer \cite{ji2024wavtokenizer} generalizes less effectively across diverse voices, languages, and recording conditions compared to EnCodec \cite{encodec} and DAC \cite{kumar2024high}. However, its semantically richer latent space may contribute to higher overall generation quality. This could explain why \textsc{Lina-Speech} outperforms larger or end-to-end models, particularly on LibriTTS. Future work in latent audio modeling should be considered to better understand and refine our architectural improvements.

\paragraph{Streamability} Although \textsc{Lina-Speech} demonstrates nearly linear time complexity within its context window, additional work is needed in order to enable seamless streaming. To achieve this, we plan to explore chunk-based text encoding and windowed cross-attention to enable fully linear, streaming synthesis. 

\paragraph{Initial-state tuning} 

In this work we introduced low-rank structured initial-state. This approach has been successful for adapting to a small amount of data samples and may benefit other modalities such as natural language. We also plan to use initial-state tuning for generating high-quality synthetic dataset.

\section{Conclusion}

In this paper, we introduce \textsc{Lina-Speech}, a novel text-to-speech (TTS) model built upon a neural codec language model. We demonstrate that Gated Linear Attention (GLA) constitutes a robust foundation for scalable TTS systems, achieving state-of-the-art performance while substantially improving inference throughput. Moreover, we propose a new technique for conditioning the model on a larger amount of audio by tuning a low-rank constrained initial state. This approach effectively addresses the limitations of fixed context length, allowing the model to handle more—and longer—conditioning audio efficiently. Our experimental results show that this method leads to significant improvements in tasks such as audiobooks narration and expressive speech generation.


\section{Impact statement}

This paper presents an approach to improving TTS through principled architecture choices and better prompting strategies.
While our work improves the quality and flexibility of voice cloning, we acknowledge potential ethical concerns, particularly regarding misuse in deepfake generation, unauthorized voice replication, and speaker identity theft. To mitigate these risks, we advocate for the responsible use of our techniques, including watermarking, speaker verification safeguards, and adherence to ethical AI deployment guidelines.
By improving controllability and fidelity in TTS, our research contributes to both scientific advancements and practical applications, with the potential to benefit individuals who rely on synthetic speech while emphasizing ethical considerations in voice cloning technology.

\bibliography{aaai2026}

\end{document}